# Hybrid organic-inorganic two-dimensional metal carbide MXenes with amido- and imido-terminated surfaces

Chenkun Zhou[1], Di Wang[1], Francisco Lagunas[2], Benjamin Atterberry[3,4], Ming Lei[5], Huicheng Hu[1], Zirui Zhou[1], Alexander S. Filatov[1], De-en Jiang[5], Aaron J. Rossini[3,4], Robert F. Klie[2], Dmitri V. Talapin[1,6*]

[1]*Department of Chemistry, James Franck Institute, and Pritzker School of Molecular Engineering, University of Chicago, Chicago, Illinois 60637, United States*

[2]*Department of Physics, University of Illinois Chicago, Chicago, Illinois 60607, United States*

[3]*US Department of Energy, Ames Laboratory, Ames, Iowa 50011, United States*

[4]*Department of Chemistry, Iowa State University, Ames, Iowa 50011, United States*

[5]*Department of Chemical and Biomolecular Engineering, Vanderbilt University, Nashville, Tennessee 37235, United States*

[6]*Center for Nanoscale Materials, Argonne National Laboratory, Argonne, Illinois 60439, United States*

*Correspondence to: dvtalapin@uchicago.edu

**Two-dimensional (2D) transition-metal carbides and nitrides (MXenes)[1, 2] show impressive performance in applications, such as supercapacitors[3], batteries[4], electromagnetic interference shielding[5, 6], or electrocatalysis[7]. These materials combine the electronic and mechanical properties of 2D inorganic crystals with chemically modifiable surfaces[8], and surface-engineered MXenes represent an ideal platform for fundamental and applied studies of interfaces in 2D functional materials. A natural step in structural engineering of MXene compounds is the development and understanding of MXenes with various organic functional groups covalently bound to inorganic 2D sheets[9, 10, 11]. Such hybrid structures have the potential to unite the tailorability of organic molecules with the unique electronic properties of inorganic 2D solids. Here, we introduce a new family of hybrid MXenes (*h*-MXenes) with amido- and imido-bonding between organic and inorganic parts. The description of *h*-MXene structure requires an intricate mix of concepts from the fields of coordination chemistry, self-assembled monolayers (SAMs) and surface science. The optical properties of *h*-MXenes reveal coherent coupling between the organic and inorganic components. *h*-MXenes also show superior stability against hydrolysis in aqueous solutions.**

MXenes are typically prepared from MAX phases, where M represents an early transition metal (e.g., Ti, Nb, V, Mo, etc.), A represents element mainly from groups 13-16 (Al, Si, etc.), and X stands for C or N[12]. MAX phases are converted to 2D MXenes by selectively etching away A-layer elements, typically using fluoride solutions[12, 13, 14]. The resulting exfoliated MXenes have a mixture of -F, -O, and -OH surface termination groups usually denoted as $T_x$. Unlike surfaces of graphene and transition-metal dichalcogenides, the basal surfaces of MXenes allow for further chemical modification with different functional groups. However, the very strong Ti-F and Ti-O bonds introduced during MAX exfoliation with fluoride reagents make post-synthetic substitutions of $T_x$ surface groups difficult. New synthetic routes, which omit fluoride by transferring the MAX etching process to a molten-salt medium, can produce MXenes with pure Cl or Br terminations[8, 15, 16, 17]. Ti-Cl and Ti-Br surface bonds are labile enough to allow the exchange of surface halogen atoms with other groups, and MXenes with oxo- ($Ti_3C_2O$), imido- ($Ti_3C_2NH$), thio- ($Ti_3C_2S$), seleno- ($Ti_3C_2Se$), or telluro- ($Ti_3C_2Te$) terminations, as well as bare MXenes ($Ti_3C_2\square_2$), can all be prepared from $Ti_3C_2Cl_2$ or $Ti_3C_2Br_2$. The surface groups can define MXene properties, such as superconductivity[8] and electrochemical energy storage capacitance[15]. Theoretical studies have predicted that surface terminations control many other physical and chemical properties of MXenes[18, 19, 20].

Here, we introduce a general approach toward hybrid organic-inorganic MXenes (*h*-MXenes) with a broad scope of organic terminal groups. *h*-MXenes can be synthesized *via* the displacement of halides from Br- or Cl-terminated MXenes by deprotonated primary organic amines (e.g., *n*-alkylamines). The amines are first deprotonated by NaH or *n*-butyllithium, and then reacted with multilayer $Ti_3C_2Br_2$ MXenes at 120 °C for two days (see Methods). The overall reaction between $Ti_3C_2Br_2$ and amine in presence of deprotonating agent NaH is as follows:

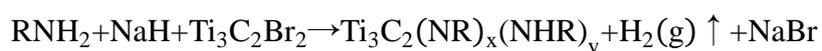

$RNH_2 + NaH + Ti_3C_2Br_2 \rightarrow Ti_3C_2(NR)_x(NHR)_y + H_2(g)\uparrow + NaBr$

The scope of studied amines is shown in Fig. 1. Complete removal of Br was confirmed by X-ray fluorescence (XRF) analysis (Extended Data Fig. 1a). Deprotonation of amines is crucial to initiate the exchange reaction on MXene surfaces; tertiary amines (e.g., triethylamine) showed no reactivity toward halide terminated MXenes (Extended Data Fig. 1c). Identical products were obtained by reacting $Ti_3C_2Br_2$ and $Ti_3C_2Cl_2$ with deprotonated butylamine (Extended Data Fig. 2a). Very strong binding of amines to $Ti_3C_2$ sheets follows from thermogravimetric analysis, which shows that, for example, the *n*-propylamine-functionalized $Ti_3C_2(pra)_{2/3}$ shows only 4% weight loss when heated from 50 °C to 400 °C (Extended Data Fig.

1d). Besides various alkylamines, this approach has been applied to diamines, aromatic amines, and poly(ethylene glycol) (PEG)-amines. Examples of the corresponding *h*-MXenes are shown in Extended Data Fig. 2b, c. The above reaction can also be applied to $Ti_2CCl_2$ and $Nb_2CCl_2$ to obtain organic-inorganic hybrid $Ti_2C$ or $Nb_2C$ MXenes, as shown in Extended Data Fig. 2d.

Powder X-ray diffraction (XRD) patterns from $Ti_3C_2$-derived *h*-MXenes with different *n*-alkyl groups are shown in Fig. 2a. All structures can be assigned to the $P6_3/mmc$ space group. The first member of this homologous series, $Ti_3C_2(NH)_x$ where $x\sim1$, can be synthesized by reacting $Ti_3C_2Br_2$ with $NaNH_2$. With an increase of the length of alkyl chain, all (0 0 0 *l*) diffraction peaks of *h*-MXenes shift to smaller 2θ angles, indicating the expansion of the spacing between $Ti_3C_2$ layers. Some diffraction peaks, e.g., the (0 0 0 10) reflection of $Ti_3C_2(pra)_{2/3}$, disappear due to symmetry-related cancellation (Extended Data Fig. 3). The lattice parameters of *h*-MXenes, calculated using the Le Bail method, show that the *c*-lattice constant linearly increases with the length of alkyl chain (Fig. 2b), which implies that the surface coverage is independent of alkyl chain length.

The *a*-lattice constant, which defines the distance between neighboring Ti atoms on the surface of $Ti_3C_2$ sheets, is practically independent of the alkyl chain length. Interestingly, the interatomic metal-metal distances at the surface of *h*-$Ti_3C_2$ MXenes (3.04 Å) are very close to those at Au (111) surface (2.88 Å), and packing of alkyl chains on these basal planes could resemble the structure of self-assembled monolayers (SAMs) of *n*-alkanethiol molecules on Au (111) surfaces[21, 22]. For the Au (111) surface, SAM grafting density is limited by the steric bulk of alkyl tails: the organic layer fills space completely with alkyl tails tilted approximately 30° from the surface normal to maximize their van der Waals interactions (Fig. 2c), whereas surface-binding head groups occupy three-fold sites of a $(\sqrt{3}\times\sqrt{3})R30°$ lattice[22]. Such packing produces a $Ti_3C_2(NR)_{2/3}$ or $Ti_3C_2(NHR)_{2/3}$ stoichiometry, which is in good agreement with the elemental analysis (Extended Data Fig. 4a).

The spacing between $Ti_3C_2$ sheets can be compared to the length of fully extended alkyl chains. The slope of the line in Fig. 2b shows that neighboring $Ti_3C_2$ sheets sandwich two layers of alkyl chains with ~1.10 Å per methylene unit (Extended Data Fig. 4b, c). Since the theoretical length of fully extended alkyl chains with all-*trans* conformations is ~1.27 Å per $CH_2$ group[21], the tilt of the alkyl chains in *h*-MXenes should be close to $\arccos(1.10\ \text{Å}/1.27\ \text{Å}) = 30°$ from the surface normal, again consistent with similar packing geometries of alkyl chains in *h*-MXenes and Au (111) alkanethiol SAMs.

Scanning electron microscopy (SEM) and atomic-resolution scanning transmission electron microscopy (STEM) were used to directly visualize the microstructure of *h*-MXenes. The expansion of interlayer spacing between $Ti_3C_2$ sheets with the introduction of amines can be clearly observed in STEM images (Fig. 2d and Extended Data Fig. 5a, b). Electron energy loss spectroscopy (STEM-EELS, Extended Data Fig. 5e) reveals the expected distribution of Ti, C, and N atoms along the *c*-axis of *h*-MXenes, and SEM-EDS mapping confirms uniform distribution of terminating groups in macroscopic MXene stacks (Fig. 2e). To visualize organic termination groups, we imaged $Ti_3C_2$(tma) containing heavy sulfur atoms using atomic-resolution ABF STEM. A double-layered arrangement of tentatively 2-thiophenemethylimido groups was observed between $Ti_3C_2$ sheets (Extended Data Fig. 6d), supporting the ordering of organic groups in *h*-MXenes.

Among various possible *h*-MXenes, those prepared from diamines, such as ethylenediamine (en) or *N*-methylethylenediamine (nmeda), are of particular interest as two nitrogen atoms can adopt different bonding motifs on Ti surface. Since the observed interlayer distance in $T_3C_2$(en) is larger than the length of extended en molecule, it appears the diamine ligands cannot bridge neighboring $Ti_3C_2$ sheets. When diamine *h*-MXene $Ti_3C_2$(nmeda) was treated with dilute hydrobromic acid, X-ray photoelectron spectroscopy (XPS) showed the formation of $-NRH_3^+$ species (Extended Data Fig. 7a, b). Meanwhile, the (0 0 0 2) peak in the XRD pattern shifted to smaller 2θ angles, indicative of expansion of the interlayer distance after the protonation (Extended Data Fig. 7c). Accordingly, we conclude that in diamine $h$-MXenes, one nitrogen atom is bonded to the surface of $Ti_3C_2$ sheet while the other remains chemically accessible for protonation.

Infrared (IR) absorption spectroscopy has been routinely used to study vibrations of molecules in SAMs bound to extended metal surfaces[22] or metal nanoparticles[23, 24]. Qualitatively, the vibrational spectra of surface-bound alkyl chains resemble spectra of corresponding molecules not bound to metal surface because the molecular vibrations, e.g., C-H stretches, are practically not affected by the metal surface. A very different behavior is observed for vibrational spectra of $h$-MXenes: the vibronic absorption bands of $n$-alkylamine exhibit asymmetric line shapes characteristic of Fano resonances[25, 26] (Fig. 3a and Supplementary Discussion 1). Such resonances emerge when quantum states with discrete spectra, such as vibrational normal modes, coherently couple with a continuum band of states, e.g., a plasmon or polariton mode[26]. The constructive or destructive interference of two paths (Fig. 3b) creates a characteristic asymmetric Fano line shape. In traditional SAMs on flat metal surfaces, or alkyl chains tethered to the surface of individual metal nanocrystals, this coupling of discrete and continuum states is too weak to develop the Fano effect. For $h$-MXenes, however, the coherent coupling of organic and inorganic components does appear to be sufficiently strong, and we suggest it may originate from plasmonic enhancement of electromagnetic field between $Ti_3C_2$ sheets[27]. For example, Fano resonances have been reported for molecular vibrations resonantly coupled to plasmonic hot spots with strong optical field enhancement[28]. From a practical view, the coupling of organic and inorganic components of $h$-MXenes can be used to facilitate charge, heat, and energy transfer across the organic-inorganic interfaces.

To investigate bonding details between the organic and inorganic components of $h$-MXenes, we employed magic angle spinning (MAS) solid-state NMR spectroscopy (ssNMR), which has proven to be a powerful tool for probing chemical bonding at MXene surfaces (Supplementary Discussion 2)[29, 30, 31, 32]. The one-dimensional $^1$H spin echo spectrum of $^{15}$N-labeled Ti$_3$C$_2$($^{15}$N-dda)$_{2/3}$ h-MXene shows an intense peak at 1 ppm associated with overlapping CH$_2$ signals from dodecyl alkane hydrogens. In addition, broader and lower-intensity $^1$H NMR signals centered at chemical shifts of 9 ppm and 20 ppm are visible (Fig. 3c). The latter is assigned to the NH hydrogen atoms of amido ligands, while the former is tentatively assigned to surface hydride based on a hypothesized pathway of amido to imido conversion (Supplementary Discussion 2). The $^1$H→$^{15}$N cross-polarized MAS (CPMAS) spin echo spectrum shows two broad $^{15}$N NMR signals with chemical shifts 30 and -28 ppm, which are assigned to imido (NR) and amido (NHR) ligands that coordinate to surface titanium atoms. These assignments are consistent with previously reported $^{15}$N chemical shifts for molecular transition metal imido and amido complexes, where the former is more positively shifted[33]. Peak fitting of the $^{15}$N CPMAS spin echo NMR spectrum suggests that ca. 58% of the nitrogen atoms are in the imido form, while 42% are in the amido form (Supplementary Discussion 2). The $^{15}$N signal assignments were further confirmed with $^1$H{$^{15}$N} indirectly detected heteronuclear correlation (idHETCOR) experiments obtained with backwards CP contact times of 0.4 ms or 8 ms to probe short- and long-range $^1$H-$^{15}$N internuclear distances, respectively (Fig. 3d). As expected, the HETCOR spectrum obtained with a 0.4 ms contact time only shows the amido $^{15}$N NMR signal centered at -28 ppm and reveals that it correlates to the $^1$H NMR signal at +20 ppm. The HETCOR spectrum obtained with 8 ms contact time shows both imido and amido $^{15}$N NMR signals. Acquisition of 1D $^1$H{$^{15}$N} idHETCOR spectra with variable backwards CP contact time allows measurement of $^1$H-$^{15}$N dipolar coupling constants and estimation of $^1$H-$^{15}$N internuclear distances (Supplementary Discussion 2)[34]. This experiment confirms that the amido hydrogen atom (δ = 20 ppm) has a 1.04 Å N-H internuclear distance and suggests that $^1$H hydrogen atoms (δ = 9 ppm) are ca. 1.9 Å from amido and imido nitrogen atoms. Finally, $^1$H-detected $^{15}$N{$^1$H} heteronuclear *J*-resolved spin echo experiments were used to confirm that the amido nitrogen makes a covalent bond to a single hydrogen atom (Fig. 3e). The *J*-resolved curve obtained by monitoring the amido $^1$H NMR signal can be fit with a cosine function, confirming there is only a single attached hydrogen atom. The fit yields a nitrogen-hydrogen scalar coupling constant ($^1J_{NH}$) of 44 Hz. A *J*-resolved curve for alkyl $^1$H NMR signal obtained with a back-CP contact time of 8 ms can be fit with two $^1J_{NH}$ values of 44 Hz and 0 Hz. The latter must correspond to imido nitrogen atoms, as these nitrogen atoms do not have a covalent bond to hydrogen. Using Green's classification of covalent bonds, the amido groups can be classified as L-µ-X type ligands and the imido groups as L-µ$^2$-X$_2$ type ligands (Fig. 3f and Supplementary Discussion 4)[35]. This classification has been widely used for nanocrystal surfaces, and proves particularly useful in describing ligand substitution reactions[36].

*h*-MXenes with an appropriate choice of surface-bound alkyl chains can be delaminated and dispersed in nonpolar solvents to form colloidal solutions (Fig. 4 and Extended Data Fig. 8). Colloidal stabilization in non-polar solvents typically requires a negative free energy of chain-solvent mixing[37], which causes the hydrocarbon chains to repel one another, thus stabilizing colloidal dispersions[37]. This approach works very well for small nanocrystals where surface curvature allows solvent molecules to efficiently penetrate between surface-bound alkyl chains. At the same time, chain-solvent mixing is known to be inefficient at flat surfaces with densely packed alkyl chains[37]. Accordingly, *h*-MXenes with only one type of alkyl chain did not show good colloidal stability in non-polar solvents (Fig. 4a). However, simultaneous incorporation of short (e.g., octyl) and long (e.g., oleyl) chains greatly improved the colloidal stability of *h*-MXenes by providing room for the free rotation of long chains and efficient interaction with solvent molecules (Fig. 4b). Such combinations of organic ligands, known to the nanocrystal community as "entropic ligands"[38], can efficiently produce colloidal dispersions of *h*-MXenes in CHCl$_3$ with high solid concentrations (Fig. 4b and Extended Data Fig. 8d, e). Raman spectroscopy showed that colloidal *h*-MXenes possess the same surface groups as their bulk counterparts (Extended Data Fig. 8g). Surface exchange of $Ti_3C_2Br_2$ with the mixture of propylamine and $PEG_{1K}$ amine resulted in $Ti_3C_2$ MXenes with PEGylated surfaces that are easily dispersed in water (Fig. 4c and Extended Data Fig. 8h).

MXenes bring many exciting opportunities, but their relatively poor stability against hydrolysis, especially in basic solutions, has been a source of legitimate concerns and an obvious area for improvement. Different strategies have been explored to stabilize MXenes[39, 40, 41, 42, 43], but the effect of surface groups on the hydrolytic stability is yet to be investigated. We performed a comparative study of the hydrolysis rates for traditional $Ti_3C_2T_x$ (T=F, OH, O) MXenes with $Ti_3C_2$(NH) as well as $Ti_3C_2$(pra)$_{2/3}$ and $Ti_3C_2$(oca)$_{2/3}$ *h*-MXenes. To minimize the effect of sample preparation conditions, we compared stability of multilayer MXenes of similar size in pure water at room temperature and at 71 °C. A combination of XRD, XPS, and Raman spectroscopy was used to monitor sample evolution, and a full description of our stability study is provided in Supplementary Discussion 5. At room temperature, $Ti_3C_2T_x$, $Ti_3C_2$(NH) and *h*-MXenes showed no obvious degradation after 35 days in air-saturated D.I. water (Supplementary Discussion 5). However, in accelerated tests at 71 °C, the $Ti_3C_2T_x$ sample showed significant amounts of anatase $TiO_2$ phase formed due to hydrolysis after 7 days, while no $TiO_2$ phase was detected in powder-XRD patterns and in Raman spectra for $Ti_3C_2$(NH) and *h*-MXene samples (Figs. 4d, e). *h*-MXenes also showed significantly improved stability in 0.01 M KOH solutions (Supplementary Discussion 5). The alkyl chains provide additional protection against hydrolysis by creating thin hydrophobic barriers, but since we observed a similar stability improvement for $Ti_3C_2$(NH), $Ti_3C_2$(pra)$_{2/3}$ and $Ti_3C_2$(oca)$_{2/3}$, it is reasonable to suggest that surface Ti atoms of amido/imido-terminated MXenes are less susceptible to nucleophilic attack by hydroxyl ions, while hydrophobic surface encapsulation is a complementary and probably secondary effect. As a word of caution, $Ti_3C_2$(NH) and *h*-MXenes, as with bulk TiN and TiC[44, 45], are not fully immune to oxidative hydrolysis in hot water. For example, XPS studies show the presence of $TiO_2$ at the surface of all MXene samples after one week of exposure to air-saturated water at 71 °C (Supplementary Discussion 5). We attribute it to slow dissolution of titanium species from MXene edges, followed by precipitation of a thin $TiO_2$ layer. This layer can be only a few nanometers thick since etching of sample surface with $Ar_{500}^+$ clusters efficiently restores the original *h*-MXene surface (Supplementary Discussion 5). Our studies suggest that amido/imido-surface chemistry generally improves MXene resistance against hydrolysis and shows that surface engineering is a viable strategy toward synthesis of functional MXenes with enhanced stability.

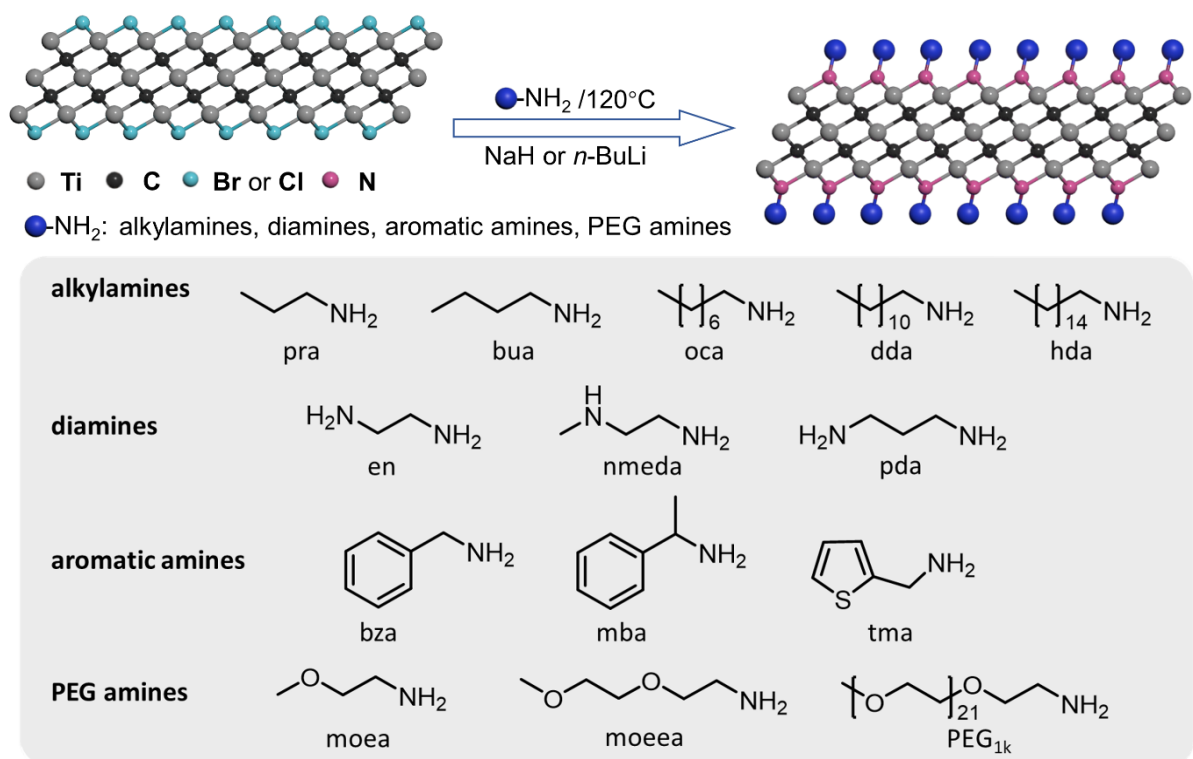

**Fig. 1|** Schematic of organic-inorganic *h*-MXene synthesis and examples of studied organic amines.

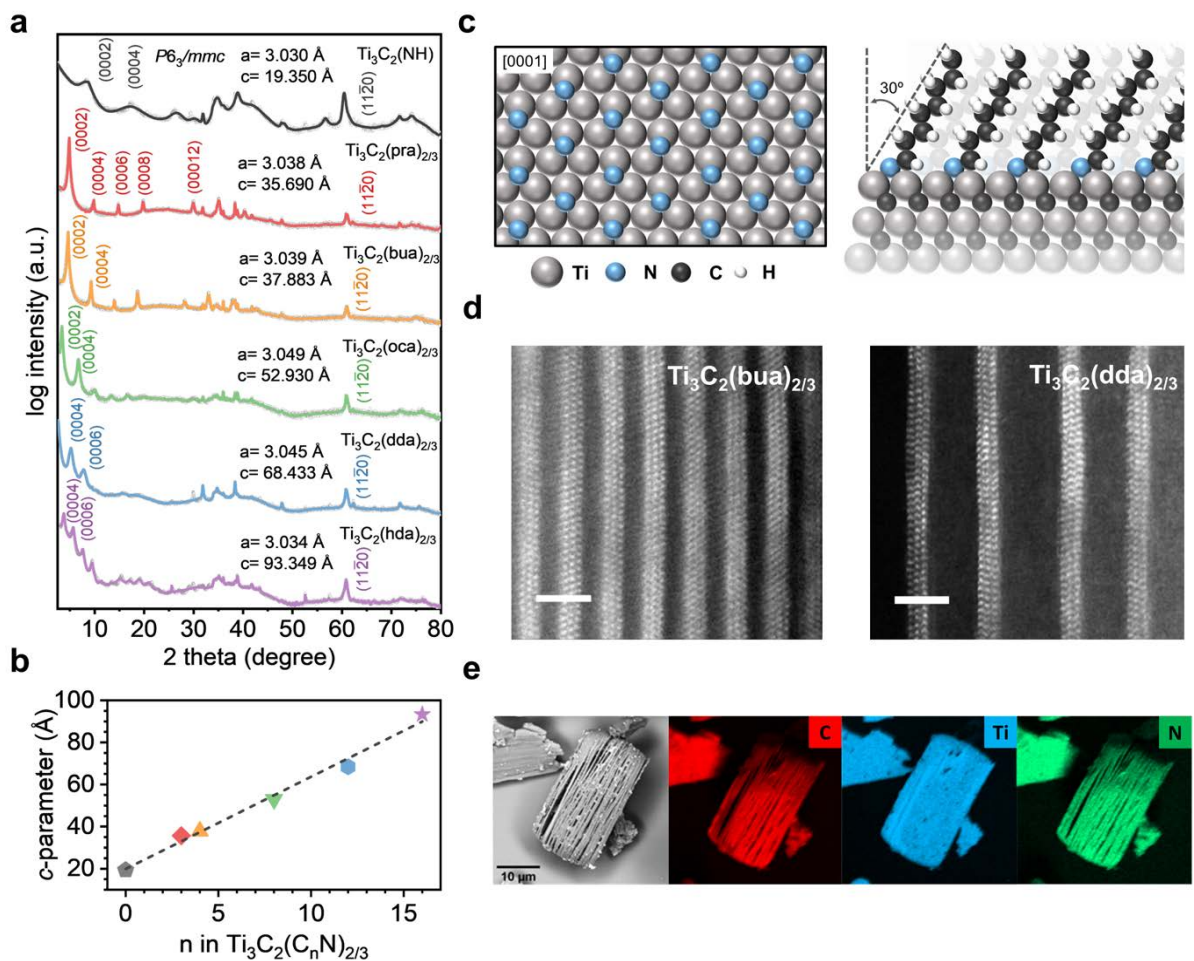

**Fig. 2| Structural characterizations of *h*-MXenes. a,** Powder XRD patterns and their Le Bail fits for Ti$_3$C$_2$(NH) and *h*-MXenes with alkylimido surface-termination groups. **b,** Changes of the lattice *c* parameter versus number of carbons in the interlayer alkyl groups. **c,** Schematic of alkylimido surface termination groups on a Ti$_3$C$_2$ sheet. **d,** LAADF-STEM images (scale bar: 2 nm) showing the lamellar structures of *h*-MXenes; image of Ti$_3$C$_2$(dda)$_{2/3}$ was acquired using Cryo-STEM method. **e,** SEM-EDS element mapping of Ti$_3$C$_2$(bua)$_{2/3}$ MXene.

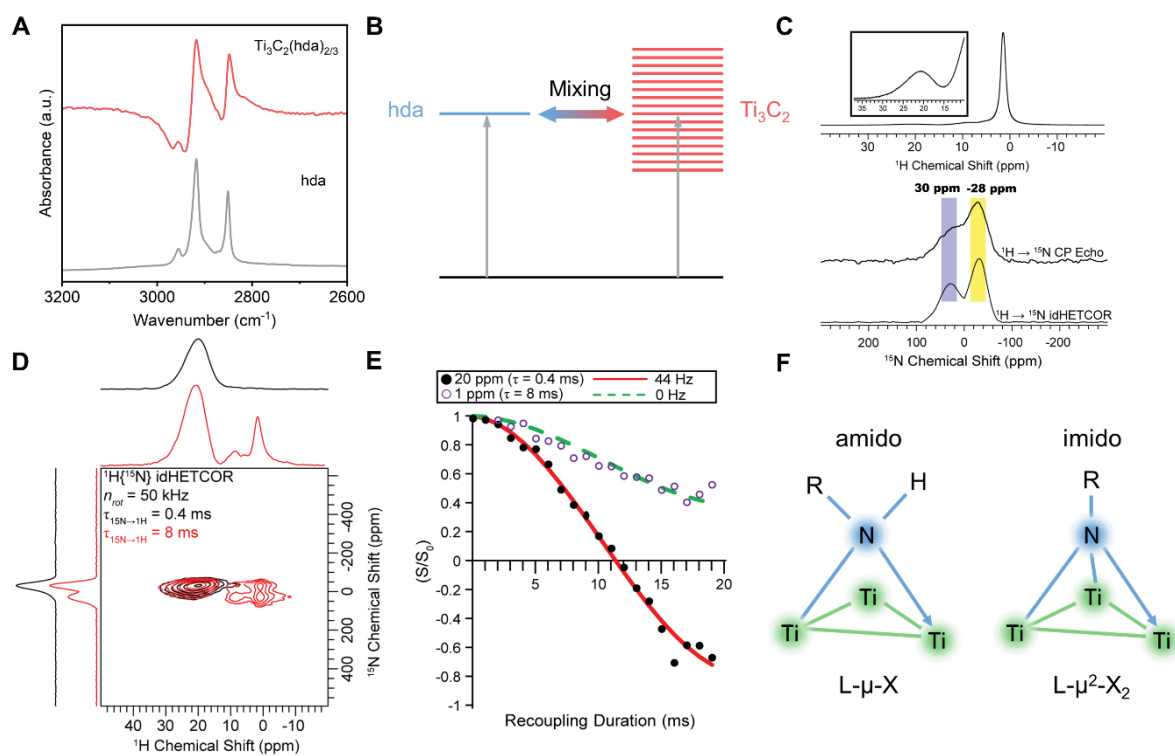

**Fig. 3| Fano effect in infrared absorption spectra and magic angle spinning (MAS) solid state NMR spectra of *h*-MXenes. a,** C-H bond stretching region of infrared absorption spectra of $Ti_3C_2(hda)_{2/3}$ showing the Fano resonant coupling compared with the absorption of neat hexadecylamine. **b,** Schematic of the Fano coupling between discrete vibrational modes of *n*-alkylamines and the continuum band of states in metallic $Ti_3C_2$ sheets. **c,** $^1H$ spin echo, $^1H→^{15}N$ CPMAS spin echo, and $^1H$-detected $^{15}N$ CPMAS spectrum of $Ti_3C_2(^{15}N\text{-dda})_{2/3}$. **d,** 2D $^1H\{^{15}N\}$ idHETCOR spectra obtained with backwards CP contact times of 0.4 ms or 8 ms to probe short- and long-range $^1H$-$^{15}N$ internuclear distances, respectively. **e,** $^1H$ detected $^{15}N\{^1H\}$ heteronuclear spin echo (*J*-resolved) curves confirming that the $^{15}N$ NMR signal at –27 ppm corresponds to an amido nitrogen (NHR), while the $^{15}N$ NMR signal at 30 ppm corresponds to a deprotonated imido nitrogen (NR). All spectra were obtained with an MAS frequency of 50 kHz. **f,** The bonding motifs between the organic and inorganic components of *h*-MXenes.

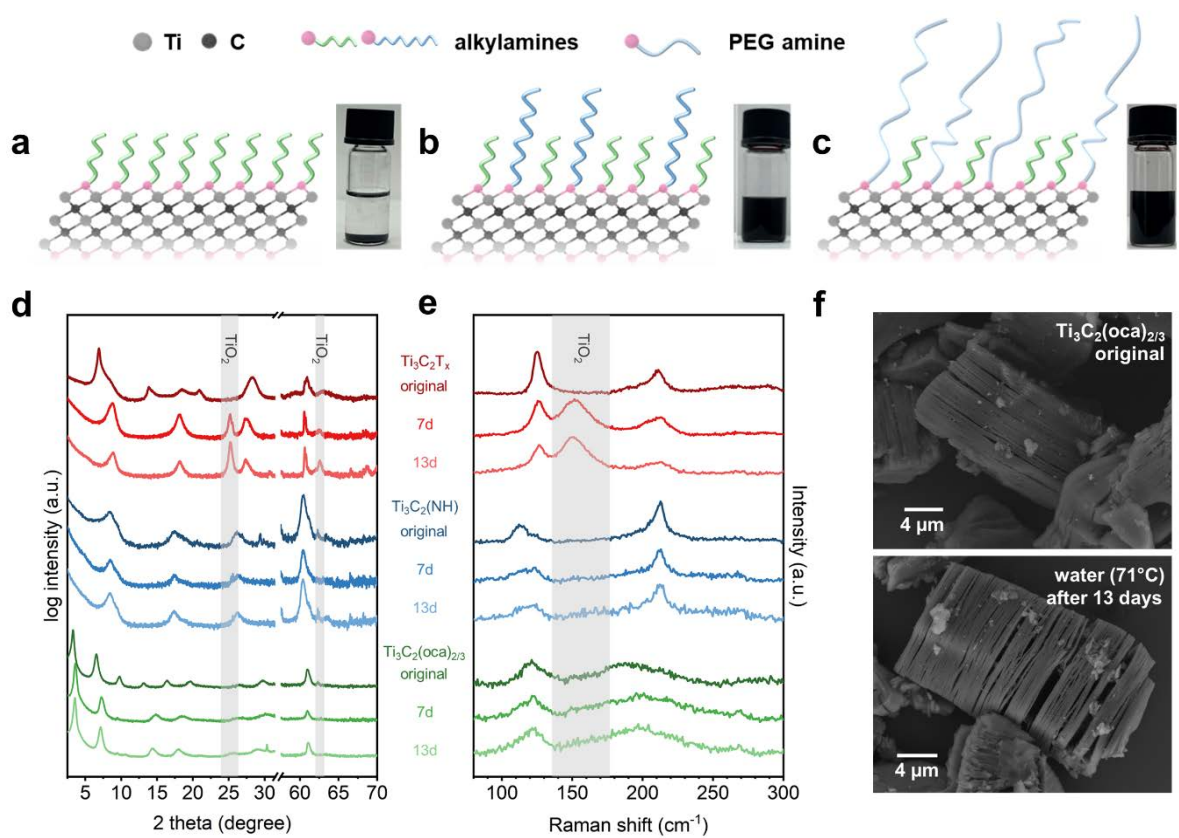

**Fig. 4| Delamination, colloidal dispersion, and hydrolytic stability of *h*-MXenes.** Schematics of delaminated *h*-MXenes with different types of surface ligands. Inset: Photographs of **a,** *d*-Ti$_3$C$_2$(oca)$_{2/3}$ in CHCl$_3$, **b,** *d*-Ti$_3$C$_2$(oca/ola) in CHCl$_3$ and **c,** *d*-Ti$_3$C$_2$(pra/PEG$_{1k}$) in water. **d,** Powder XRD patterns and **e,** Raman spectra of Ti$_3$C$_2$T$_x$, Ti$_3$C$_2$(NH), and Ti$_3$C$_2$(oca)$_{2/3}$ after exposure to 71 °C pure water for 7 days and 13 days. **f,** SEM images of original Ti$_3$C$_2$(oca)$_{2/3}$ and after exposure to 71 °C pure water for 13 days.

# References

1. Anasori, B., Lukatskaya, M.R., Gogotsi, Y. 2D metal carbides and nitrides (MXenes) for energy storage. *Nat. Rev. Mater.* **2**, 1-17 (2017).

2. VahidMohammadi, A., Rosen, J., Gogotsi, Y. The world of two-dimensional carbides and nitrides (MXenes). *Science* **372**, eabf1581 (2021).

3. Xia, Y., Mathis, T.S., Zhao, M.-Q., Anasori, B., Dang, A., Zhou, Z., *et al.* Thickness-independent capacitance of vertically aligned liquid-crystalline MXenes. *Nature* **557**, 409-412 (2018).

4. Zhang, C.J., Park, S.-H., Seral-Ascaso, A., Barwich, S., McEvoy, N., Boland, C.S., *et al.* High capacity silicon anodes enabled by MXene viscous aqueous ink. *Nat. Commun.* **10**, 1-9 (2019).

5. Iqbal, A., Shahzad, F., Hantanasirisakul, K., Kim, M.-K., Kwon, J., Hong, J., *et al.* Anomalous absorption of electromagnetic waves by 2D transition metal carbonitride $Ti_3CNT_x$ (MXene). *Science* **369**, 446-450 (2020).

6. Shahzad, F., Alhabeb, M., Hatter, C.B., Anasori, B., Hong, S.M., Koo, C.M., *et al.* Electromagnetic interference shielding with 2D transition metal carbides (MXenes). *Science* **353**, 1137-1140 (2016).

7. Zhang, J., Zhao, Y., Guo, X., Chen, C., Dong, C.-L., Liu, R.-S., *et al.* Single platinum atoms immobilized on an MXene as an efficient catalyst for the hydrogen evolution reaction. *Nat. Catal.* **1**, 985-992 (2018).

8. Kamysbayev, V., Filatov, A.S., Hu, H., Rui, X., Lagunas, F., Wang, D., *et al.* Covalent surface modifications and superconductivity of two-dimensional metal carbide MXenes. *Science* **369**, 979-983 (2020).

9. Kim, D., Ko, T.Y., Kim, H., Lee, G.H., Cho, S., Koo, C.M. Nonpolar organic dispersion of 2D $Ti_3C_2T_x$ MXene flakes via simultaneous interfacial chemical grafting and phase transfer method. *ACS Nano* **13**, 13818-13828 (2019).

10. Mayorga-Burrezo, P., Muñoz, J., Zaoralová, D., Otyepka, M., Pumera, M. Multiresponsive 2D $Ti_3C_2T_x$ MXene via implanting molecular properties. *ACS Nano* **15**, 10067-10075 (2021).

11. Park, G.S., Ho, D.H., Lyu, B., Jeon, S., Ryu, D.Y., Kim, D.W., *et al.* Comb-type polymer-hybridized MXene nanosheets dispersible in arbitrary polar, nonpolar, and ionic solvents. *Sci. Adv.* **8**, eabl5299 (2022).

12. Naguib, M., Kurtoglu, M., Presser, V., Lu, J., Niu, J., Heon, M., *et al.* Two-dimensional nanocrystals produced by exfoliation of $Ti_3AlC_2$. *Adv. Mater.* **23**, 4248-4253 (2011).

13. Ghidiu, M., Lukatskaya, M.R., Zhao, M.-Q., Gogotsi, Y., Barsoum, M.W. Conductive two-dimensional titanium carbide 'clay' with high volumetric capacitance. *Nature* **516**,

78-81 (2014).

14. Feng, A., Yu, Y., Jiang, F., Wang, Y., Mi, L., Yu, Y., *et al.* Fabrication and thermal stability of $NH_4HF_2$-etched $Ti_3C_2$ MXene. *Ceram. Int.* **43**, 6322-6328 (2017).

15. Li, Y., Shao, H., Lin, Z., Lu, J., Liu, L., Duployer, B., *et al.* A general Lewis acidic etching route for preparing MXenes with enhanced electrochemical performance in non-aqueous electrolyte. *Nat. Mater.* **19**, 894-899 (2020).

16. Li, M., Lu, J., Luo, K., Li, Y., Chang, K., Chen, K., *et al.* Element replacement approach by reaction with Lewis acidic molten salts to synthesize nanolaminated MAX phases and MXenes. *J. Am. Chem. Soc.* **141**, 4730-4737 (2019).

17. Li, M., Li, X., Qin, G., Luo, K., Lu, J., Li, Y., *et al.* Halogenated $Ti_3C_2$ MXenes with electrochemically active terminals for high-performance zinc ion batteries. *ACS Nano* **15**, 1077-1085 (2021).

18. Khazaei, M., Arai, M., Sasaki, T., Chung, C.-Y., Venkataramanan, N.S., Estili, M., *et al.* Novel electronic and magnetic properties of two-dimensional transition metal carbides and nitrides. *Adv. Funct. Mater.* **23**, 2185-2192 (2013).

19. Liu, Y., Xiao, H., Goddard, W.A. Schottky-barrier-free contacts with two-dimensional semiconductors by surface-engineered MXenes. *J. Am. Chem. Soc.* **138**, 15853-15856 (2016).

20. Si, C., Zhou, J., Sun, Z. Half-metallic ferromagnetism and surface functionalization-induced metal–insulator transition in graphene-like two-dimensional $Cr_2C$ crystals. *ACS Appl. Mater. Interfaces* **7**, 17510-17515 (2015).

21. Bain, C.D., Troughton, E.B., Tao, Y.T., Evall, J., Whitesides, G.M., Nuzzo, R.G. Formation of monolayer films by the spontaneous assembly of organic thiols from solution onto gold. *J. Am. Chem. Soc.* **111**, 321-335 (1989).

22. Dubois, L.H., Nuzzo, R.G. Synthesis, structure, and properties of model organic surfaces. *Annu. Rev. Phys. Chem.* **43**, 437-463 (1992).

23. Badia, A., Cuccia, L., Demers, L., Morin, F., Lennox, R.B. Structure and dynamics in alkanethiolate monolayers self-assembled on Gold nanoparticles: a DSC, FT-IR, and deuterium NMR study. *J. Am. Chem. Soc.* **119**, 2682-2692 (1997).

24. Hostetler, M.J., Stokes, J.J., Murray, R.W. Infrared spectroscopy of three-dimensional self-assembled monolayers: N-alkanethiolate monolayers on Gold cluster compounds. *Langmuir* **12**, 3604-3612 (1996).

25. Luk'yanchuk, B., Zheludev, N.I., Maier, S.A., Halas, N.J., Nordlander, P., Giessen, H., *et al.* The Fano resonance in plasmonic nanostructures and metamaterials. *Nat. Mater.* **9**, 707-715 (2010).

26. Miroshnichenko, A.E., Flach, S., Kivshar, Y.S. Fano resonances in nanoscale structures. *Rev. Mod. Phys.* **82**, 2257 (2010).

27. El-Demellawi, J.K., Lopatin, S., Yin, J., Mohammed, O.F., Alshareef, H.N. Tunable multipolar surface plasmons in 2D $Ti_3C_2T_x$ MXene flakes. *ACS Nano* **12**, 8485-8493 (2018).

28. Agrawal, A., Singh, A., Yazdi, S., Singh, A., Ong, G.K., Bustillo, K.*, et al.* Resonant coupling between molecular vibrations and localized surface plasmon resonance of faceted metal oxide nanocrystals. *Nano Lett.* **17**, 2611-2620 (2017).

29. Griffith, K.J., Hope, M.A., Reeves, P.J., Anayee, M., Gogotsi, Y., Grey, C.P. Bulk and surface chemistry of the niobium MAX and MXene phases from multinuclear solid-state NMR spectroscopy. *J. Am. Chem. Soc.* **142**, 18924-18935 (2020).

30. Harris, K.J., Bugnet, M., Naguib, M., Barsoum, M.W., Goward, G.R. Direct measurement of surface termination groups and their connectivity in the 2D MXene $V_2CT_x$ using NMR spectroscopy. *J. Phys. Chem. C* **119**, 13713-13720 (2015).

31. Hope, M.A., Forse, A.C., Griffith, K.J., Lukatskaya, M.R., Ghidiu, M., Gogotsi, Y.*, et al.* NMR reveals the surface functionalisation of $Ti_3C_2$ MXene. *Phys. Chem. Chem. Phys.* **18**, 5099-5102 (2016).

32. Kobayashi, T., Sun, Y., Prenger, K., Jiang, D.-e., Naguib, M., Pruski, M. Nature of terminating hydroxyl groups and intercalating water in $Ti_3C_2T_x$ MXenes: a study by $^1H$ solid-state NMR and DFT calculations. *J. Phys. Chem. C* **124**, 13649-13655 (2020).

33. Beaumier, E.P., Billow, B.S., Singh, A.K., Biros, S.M., Odom, A.L. A complex with nitrogen single, double, and triple bonds to the same chromium atom: synthesis, structure, and reactivity. *Chem. Sci.* **7**, 2532-2536 (2016).

34. Paluch, P., Pawlak, T., Amoureux, J.-P., Potrzebowski, M.J. Simple and accurate determination of X–H distances under ultra-fast MAS NMR. *J. Magn. Reson.* **233**, 56-63 (2013).

35. Green, M.L.H. A new approach to the formal classification of covalent compounds of the elements. *J. Organomet. Chem.* **500**, 127-148 (1995).

36. Anderson, N.C., Hendricks, M.P., Choi, J.J., Owen, J.S. Ligand exchange and the stoichiometry of metal chalcogenide nanocrystals: spectroscopic observation of facile metal-carboxylate displacement and binding. *J. Am. Chem. Soc.* **135**, 18536-18548 (2013).

37. Israelachvili, J.N. *Intermolecular and surface forces*. Academic press, 2011.

38. Pang, Z., Zhang, J., Cao, W., Kong, X., Peng, X. Partitioning surface ligands on nanocrystals for maximal solubility. *Nat. Commun.* **10**, 2454 (2019).

39. Zhang, J., Kong, N., Hegh, D., Usman, K.A.S., Guan, G., Qin, S.*, et al.* Freezing titanium carbide aqueous dispersions for ultra-long-term storage. *ACS Appl. Mater. Interfaces* **12**, 34032-34040 (2020).

40. Zhang, C.J., Pinilla, S., McEvoy, N., Cullen, C.P., Anasori, B., Long, E., *et al.* Oxidation stability of colloidal two-dimensional titanium carbides (MXenes). *Chem. Mater.* **29**, 4848-4856 (2017).

41. Natu, V., Hart, J.L., Sokol, M., Chiang, H., Taheri, M.L., Barsoum, M.W. Edge capping of 2D-MXene sheets with polyanionic salts to mitigate oxidation in aqueous colloidal suspensions. *Angew. Chem.* **131**, 12785-12790 (2019).

42. Mathis, T.S., Maleski, K., Goad, A., Sarycheva, A., Anayee, M., Foucher, A.C., *et al.* Modified MAX phase synthesis for environmentally stable and highly conductive $Ti_3C_2$ MXene. *ACS Nano* **15**, 6420-6429 (2021).

43. Wang, X., Wang, Z., Qiu, J. Stabilizing MXene by hydration chemistry in aqueous solution. *Angew. Chem. Int. Ed.* **60**, 26587-26591 (2021).

44. Avgustinik, A.I., Drozdetskaya, G.V., Ordan'yan, S.S. Reaction of titanium carbide with water. *Powder Metall. Met. Ceram.* **6**, 470-473 (1967).

45. Vasilos, T., Kingery, W. Note of properties of aqueous suspensions of TiC and TiN. *J. Phys. Chem.* **58**, 486-488 (1954).